\documentclass[twocolumn,superscriptaddress,preprintnumbers,amsmath,amssymb,prl]{revtex4}

\newcommand{\IMSS}{Muon Science Laboratory and Condensed Matter Research Center, Institute of Materials Structure Science, High Energy Accelerator Research Organization (KEK-IMSS), Tsukuba, Ibaraki 305-0801, Japan}
\newcommand{\Sokendai}{Department of Materials Structure Science, The Graduate University for Advanced Studies (Sokendai), Tsukuba, Ibaraki 305-0801, Japan}
\newcommand{\MSLTokyo}{Materials and Structures Laboratory, Tokyo Institute of Technology, Yokohama, Kanagawa 226-8503, Japan}
\newcommand{\MCES}{Materials Research Center for Element Strategy, Tokyo Institute of Technology (MCES), Yokohama, Kanagawa 226-8503, Japan}
\newcommand{\NIMS}{National Institute for Materials Sciences (NIMS), 1-1, Namiki, Tsukuba, Ibaraki 305-0044, Japan}

\newcommand{\Magazine}[4]{#1 {\bf #2}, #3 (#4).}
\newcommand{\JACS}{J. Am. Chem. Soc.\xspace}

\newcommand{\PRB}{Phys.~Rev. B\xspace}
\newcommand{\PRX}{Phys.~Rev. X\xspace}
\newcommand{\JPCC}{J.~Phys. Chem. C\xspace}

\newcommand{\etal}{{\it et}~{\it al}.\xspace}
\usepackage{txfonts}
\usepackage[dvipdfmx]{graphicx}
\usepackage{dcolumn} 
\usepackage{bm} 
\usepackage{xspace}
\usepackage{ulem} 
\usepackage[hypertex,colorlinks=true,linkcolor=black,citecolor=blue,filecolor=blue,urlcolor=blue,setpagesize=false,nesting=true]{hyperref}
\begin{document}

\title{Electronic correlation in the quasi-two-dimensional electride Y$_2$C}

\author{M.~Hiraishi}
\affiliation{\IMSS}
\author{K.~M.~Kojima}
\affiliation{\IMSS}\affiliation{\Sokendai}
\author{I.~Yamauchi}
\thanks{Present address: Department of Physics, Graduate School of Science and Engineering, Saga University, Saga 840-8502, Japan}
\affiliation{\IMSS}
\author{H.~Okabe}
\affiliation{\IMSS}\affiliation{\Sokendai}
\author{S.~Takeshita}
\affiliation{\IMSS}
\author{A.~Koda}
\affiliation{\IMSS}\affiliation{\Sokendai}
\author{R.~Kadono}
\affiliation{\IMSS}\affiliation{\Sokendai}
\author{X.~Zhang}
\thanks{Present address: State Key Laboratory of Information Photonics and Optical Communications and School of Science, Beijing University of Posts and Telecommunications, Beijing 100876, China}
\affiliation{\MSLTokyo}
\author{S.~Matsuishi}
\affiliation{\MCES}
\author{H.~Hosono}
\affiliation{\MSLTokyo}\affiliation{\MCES}
\author{K.~Hirata}
\affiliation{\NIMS}
\author{S.~Otani}
\affiliation{\NIMS}
\author{N.~Ohashi}
\affiliation{\NIMS}

\begin{abstract}
 Magnetic properties of the electride compound Y$_2$C were investigated by muon spin rotation and magnetic susceptibility on two samples with different form (poly- and single-crystalline), to examine the theoretically-predicted Stoner ferromagnetism for the electride bands. There was no evidence of static magnetic order in both samples even at temperatures down to 0.024~K. For the poly-crystalline sample, the presence of a paramagnetic moment at Y sites was inferred from the Curie-Weiss behavior of the muon Knight shift and susceptibility, whereas no such tendency was observed in the single-crystalline sample. These observations suggest that the electronic ground state of Y$_2$C is at the limit between weak-to-strong electronic correlation, where onsite Coulomb repulsion is sensitive to a local modulation of the electronic state or a shift in the Fermi level due to the presence of defects/impurities.
\end{abstract}



\maketitle
Electrides are a class of materials in which electrons serve as anions (without atomic nuclei) in the positively charged lattice framework sustained by covalent bonds \cite{JLDye_2003,JLDye_2009}.
In view of their promising properties such as high electrical conductivity, low work function, and significant catalytic activity in their ideal form, electrides are drawing much attention from the research community.
However, most of the reported electrides are not stable under ambient atmosphere~\cite{JLDye_2009,Ichimura_JACS2002}, which leads to the difficulty in the development of possible applications.
The first non-aerophobic electride compound that paved the way to various applications was mayenite, [Ca$_{24}$Al$_{28}$O$_{64}$]$^{4+}$4e$^-$ (C12A7:e$^-$), which was reported in 2003 by Matsuishi \etal\cite{{Matsuishi_2003}}.
Mayenite has positively charged nano-sized Ca-Al-O cages and endohedral electrons that maintain charge neutrality of the unit cell.
The electrons at the cage center move to an empty neighboring cage by quantum tunneling, thus contributing to the high electric conductivity.
Despite its low work function ($\sim$2.4~eV, comparable to that of alkali-metals), C12A7:e$^-$ is stable in ambient environment~\cite{Toda_AdvMater2007} and can therefore be used in a wide range of applications~\cite{Kim_JACC2007,Buchammagari_OrgLett2007,Ruszak_CatalLett2008,Adachi_MaterSciEngB2009,Kitano_NatChem2012,Toda_NatCommun2013,Kitano_NatCommun2015}.
\par
Recently, a layered nitride compound with the formula Ca$_2$N was reported to be an electride~\cite{Fang_2000,Lee_Ca2N}, in which, based on bulk property measurements as well as on a recent ARPES study in combination with density functional theory (DFT), it was proposed that electrons were extended in two-dimension (2D) over the interlameller space between the calcium layers~\cite{Oh_2016}.
Since the dimensionality of the anion electrons plays an important role in determining the bulk electronic properties of the material, 2D electrides are currently under spotlight in materials science research.
\par
\begin{figure}[!b]
  \centering
  \includegraphics[width=0.8\linewidth]{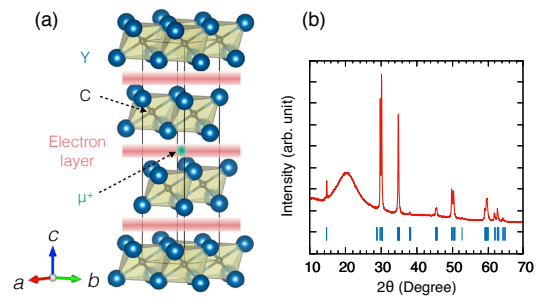}
  \caption{(Color online) (a) Crystal structure of Y$_2$C. Blue and black symbols represent Y and C atoms and the region hatched in pink shows the 2D-electron layer. Muon site predicted from the VASP calculation is shown by the green symbol. (b): Powder X-ray diffraction pattern of poly-crystalline Y$_2$C sample used in the present study. Vertical blue marks represent Bragg reflections assuming $R\bar{3}m$ space group. The halo peak near 20$^{\circ}$ originates from the sample cell.}
  \label{Fig1}
\end{figure}

Yttrium carbide Y$_2$C is one such compound, which has attracted much attention in recent time, as it is isostructural to Ca$_2$N~\cite{Inoshita_PRX,Zhang_Y2C}.
The lattice structure of these two compounds [shown in Fig.~\ref{Fig1}(a)] belongs to the space group $R{\bar 3}m$ with the lattice constant for the $a$ $(c)$-axis being 3.6164 (17.9651)~\AA\ for Y$_2$C. A DFT calculation suggests 2D electride features similar to Ca$_2$N~\cite{Zhang_Y2C} for this compound, where the electron band near the Fermi energy consists of Y 4$d$ orbitals and the 2D electronic states are present between the Y layers.
While these predicted band structures~\cite{Lee_Ca2N,Zhang_Y2C,Inoshita_theory} have been confirmed by angle-resolved photoemission spectroscopy (ARPES)~\cite{Oh_2016,Horiba_ARPES}, the interesting possibility of ferromagnetic instability associated with electride bands, which has been suggested for Y$_2$C~\cite{Inoshita_theory} is yet to be observed experimentally.
\par
In this communication, we report on the investigation of the magnetic properties of Y$_2$C using muon spin rotation ($\mu$SR) experiment on poly- (pc) and single-crystalline (sc) samples, which show different behavior in their uniform susceptibilities ($\chi$).
While other spin probes including $^{89}$Y NMR provide information for on-site spins, implanted muons can be more sensitive to the spin polarization of interlayer electrons.
From $\mu$SR measurements under zero external field (ZF-$\mu$SR), it was inferred that both samples showed no sign of magnetic order (including antiferromagnetism) at low temperatures down to 0.024~K.
Muon Knight shift measurements indicated that the observed Curie-Weiss behavior of pc $\chi$ was caused by localized moments ($\sim$1$\mu_{\rm B}$, with $\mu_{\rm B}$ being the Bohr magneton) at the Y sites. This result points to the important role of onsite electronic correlation in understanding the electride band and that, even a slight modulation of the local electronic state or a Fermi level shift due to the presence of carbon defects (often inherent) and/or impurities, causes local moments at the Y sites to appear. In contrast, the sc sample exhibited a temperature-independent negative Knight shift, much greater than that expected by the Pauli paramagnetism, suggesting that hitherto unknown contributions from orbital diamagnetism specific to the electride band may be present.
\par
Poly- and single-crystalline samples were synthesized by arc melting and floating-zone method, respectively.
The details of crystal growth and characterization of the samples are reported elsewhere~\cite{Zhang_Y2C,Otani_synthesis}.
Figure~\ref{Fig1}(b) shows the powder X-ray diffraction pattern measured at room temperature for the pc sample.
The diffraction pattern is in perfect accordance with the standard reported structure and lattice constants, indicating that the pc sample consists mostly of a single-phase. While measurements of $\chi$ show no indication of bulk ferromagnetism in both pc and sc samples, a contrasting behavior with respect to the Curie-Weiss law (see below) is observed for the two samples.

\par
Conventional $\mu$SR measurements were performed at TRIUMF, Canada, where the time-dependent ZF-$\mu$SR spectra [positron decay asymmetry, $\mathcal{A}_z(t)$] were obtained using the Lampf spectrometer for the temperature range 2--300 K and the DR spectrometer for the 0.024--4~K range. In addition, Knight shift measurements were performed under a high transverse field of $B_{\rm ext}=6$~T (HTF) and the HiTime spectrometer was used to monitor the spin precession [$\mathcal{A}_x(t)$].
A 100~\% spin-polarized muon ($\mu^+$) beam was implanted to the samples with the initial spin polarization parallel (ZF) or perpendicular (HTF) to the beam momentum direction $\hat{z}$.
The samples were loaded to cryostat under He gas atmosphere to avoid degradation, and it was confirmed after $\mu$SR measurements that their appearance was unchanged.
For the HTF-$\mu$SR measurements on sc sample, $B_{\rm ext}$ was parallel to the $c$-axis. The relative uncertainty in $B_{\rm ext}$ at the sample position ($8\times8$~mm, thickness $\sim$1~mm) was controlled to within $\sim$$\pm2$~ppm.

\begin{figure}[!t]
  \centering
  \includegraphics[width=\linewidth]{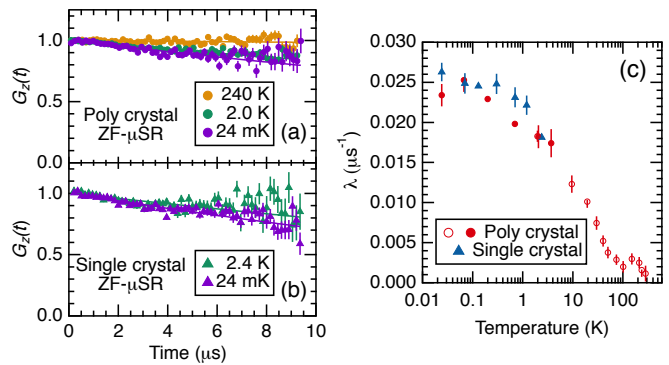}
  \caption{(Color online) Typical examples of ZF-$\mu$SR spectra for (a) poly and (b) single-crystalline samples. Solid curves are the best fits as described in the text. (c) Temperature dependence of muon depolarization rate $\lambda$ deduced from curve fit analysis. Open (closed) symbols represent data obtained by using the Lampf (DR) spectrometer.}
  \label{Fig2}
\end{figure}

Figure~\ref{Fig2} (a) and (b) show typical examples of normalized ZF-$\mu$SR time spectra, where all the spectra exhibit a slow exponential-like damping and irrespective of sample quality, little variation is observed with temperature.
This result demonstrates the absence of static magnetic order (antiferromagnetism in particular, which is often overlooked by $\chi$) over the temperature range investigated (0.024--300 K), which is in accordance with the paramagnetic behavior of $\chi$.
The spectra were then analyzed by curve fitting, assuming the following function,
\begin{align}
 \mathcal{A}_z(t) = \mathcal{A}_0G_z(t)&=\mathcal{A}_{\rm s}e^{-\lambda t}+\mathcal{A}_{\rm BG},
  \label{Eq_ZF}
\end{align}
where $\mathcal{A}_0$ is the total decay asymmetry at $t=0$, $\lambda$ is the depolarization rate due to random local fields arising from paramagnetic spins, $\mathcal{A}_s$ and $\mathcal{A}_{\rm BG}$ represent the partial asymmetry of signals from muons stopped in the sample and $T$-independent background signal due to the sample holder.
The temperature dependence of $\lambda$ is shown in Fig.~\ref{Fig2}~(c), which can be qualitatively understood to be due to the slowing down of paramagnetic fluctuation in both samples.

\begin{figure}[t]
  \centering
  \includegraphics[width=\linewidth]{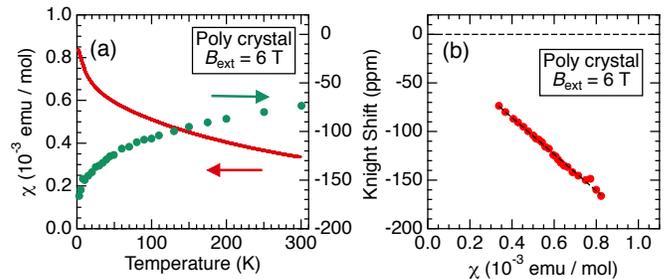}
  \caption{(Color online) (a) Temperature dependence of the magnetic susceptibility $\chi=M/H$ (left axis) and muon Knight shift $K_\mu$ (right axis) for the poly-crystalline sample measured under $B_{\rm ext}=6$~T. (b): Clogston-Jaccarino plot ($K_\mu$-$\chi$ plot) for the pc sample. Dashed line represents the best fit obtained by the linear function and its gradient corresponds to hyperfine-coupling constant $A_\mu$.}
  \label{Fig3}
\end{figure}

\begin{figure}[!t]
  \centering
  \includegraphics[width=0.6\linewidth]{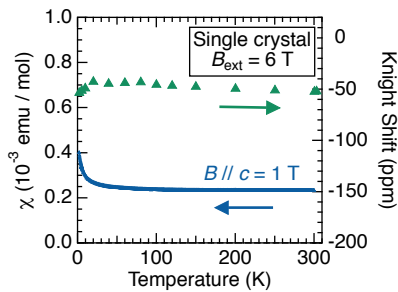}
  \caption{(Color online) Temperature dependence of molar magnetic susceptibility $\chi=M/H$ measured under 1~T (left axis) and muon Knight shift $K_\mu$ under $B_{\rm ext}=6$~T (right axis) for single-crystalline Y$_2$C. The external magnetic field was parallel to the $c$-axis.}
  \label{Fig4}
\end{figure}

The corresponding temperature ($T$) dependence of $\chi=M/H$ for the two samples are shown in the left axis of Fig.~\ref{Fig3}(a) and Fig.~\ref{Fig4}. Here, it is seen that for the pc sample, $\chi$ exhibits a divergent behavior with decreasing temperature with an enhanced gradient below $\sim$20~K.
In contrast, $\chi$ for the sc sample is mostly independent of temperature, suggesting the predominant contribution of Pauli paramagnetism.
We note that previous measurement on the same sample batch has shown that the magnitude of anisotropy, evaluated by crystalline orientation dependence of $\chi$, was very small~\cite{Otani_synthesis}.
For both the samples, the $\chi$-$T$ curve is well reproduced when using a modified Curie-Weiss law in the form $\chi(T)=\chi_0+C/(T-\Theta)$, and the parameters deduced from least-square fits are listed in Table~\ref{table}.
The large negative $\Theta$ in the pc sample indicates antiferromagnetic correlation between the local moments, whereas $\Theta\simeq0$ in the sc sample, indicating that paramagnetism arising from impurities has the dominant contribution to $\chi$ in this material.
Assuming that the local moments reside on the Y site, the effective magnetic moment ($\mu_{\rm eff}$) is deduced, to be 0.564(2)--0.605(2)$\mu_{\rm B}$/Y for the pc sample and 0.0427(2)$\mu_{\rm B}$/Y for the sc sample. While these values are comparable to those reported previously~\cite{Zhang_Y2C, Otani_synthesis}, they are in marked contrast with
the recently reported values on a sc sample that also exhibited significant magnetic anisotropy with the $c$-axis as the easy axis~\cite{Park_2017JACS}.

\begin{table}[t]
  \caption{Parameters obtained from curve fits for $\chi(T)$ using modified-Curie-Weiss law (see text). Temperature ranges used for fitting were 50--300~K and 1.8--300~K for the poly- and single-crystalline samples, respectively.}
  \centering
  \vspace{1mm} {
    \renewcommand\arraystretch{1.3}
    \tabcolsep=3mm
    \begin{tabular}{lccc}
      \hline\hline
      & $\chi_0$ [emu / mol] & $\Theta$~[K] & $\mu_{\rm eff}$ [$\mu_{\rm B}$ / Y] \\\hline
      Poly (0.4~T) & 2.236(5)$\times10^{-4}$ & $-$242.3(5) & 0.564(2)\\
      Poly (6~T) & 1.69(4)$\times10^{-5}$ & $-$264.2(4) & 0.605(2)\\
      Single (1~T) & 2.28(4)$\times10^{-4}$ & $-$3.41(3) & 0.0427(2)\\
      \hline\hline
    \end{tabular}
  }
  \label{table}
\end{table}
\par
To investigate the origin of the Curie-Weiss/Pauli paramagnetic behavior inferred from $\chi$, we performed high precision muon Knight shift measurements for the two samples for $2\le T \le300$ K.
Time spectra were analyzed by curve fitting using the following function, 
\begin{align*}
  \mathcal{A}_x(t)=\mathcal{A}_0G_x(t)=\mathcal{A}_{\rm s}e^{-\lambda t}\cos(\omega t+\phi),
\end{align*}
where $\omega=\gamma_\mu B_{\rm local}$, is the angular frequency of spin precession for muons stopped in the sample, $B_{\rm local}$ is the internal magnetic field at the muon site, $\gamma_\mu (=2\pi\times135.53$~[MHz/T]) is the muon gyromagnetic ratio, and $\phi$ is the initial phase of rotation
(We note that $\mathcal{A}_{\rm BG}\simeq 0$ for this experimental condition).
The Knight shift $K_\mu$ is then deduced from the following equations,
\begin{align*}
  K_{\mu}&=\frac{\omega-\gamma_{\mu}B_{\rm ext}}{\gamma_{\mu}B_{\rm ext}}-K_{\rm D}=\frac{A_\mu}{N_{\rm A}\mu_{\rm B}}\chi_{\rm loc},\\
  K_{\rm D}&=\left(4/3\pi-N\right)\chi,
\end{align*}
where $A_\mu$ is the muon hyperfine parameter, $N_{\rm A}$ is the Avogadro number, $\chi_{\rm loc}$ is the local susceptibility probed by the muon, $K_{\rm D}$ is the correction term for Lorentz and demagnetization fields with $N$ ($\simeq$4$\pi$) being the demagnetization factor, and $\chi$ is the uniform susceptibility.
$B_{\rm ext}$ was simultaneously monitored from the precession frequency of muons stopped in the sample holder placed under the sample that also served as a muon ``veto'' counter (made out of CaCO$_3$).
This veto counter enables to an independent recoding of the signals coming from sample and the holder.

\par
The temperature dependence of $K_{\mu}$ for the pc sample is shown in the right axis of Fig.~\ref{Fig3}(a), where a behavior qualitatively similar to that of $\chi(T)$, but with the opposite sign, is seen. The correlation of $K_\mu$ versus $\chi$ shown in Fig.~\ref{Fig3}~(b) ($K_\mu$-$\chi$ plot) exhibits a straight line, which clearly shows that the local susceptibility probed by the muon is proportional to $\chi(T)$. This demonstrates that the Curie-Weiss behavior of $\chi$ is not due to dilute impurities, but is intrinsic to the electronic state of the pc sample.
The corresponding hyperfine parameter is deduced to be $A_\mu=-52.6(3)$ [mT$/\mu_{\rm B}$/Y].

According to Hartree potential calculation for the interstitial muon using the Vienna {\it ab initio} Simulation Package (VASP)~\cite{VASP}, the muon site is estimated to be at the inter-layer 3$c$ site (0, 0, 0.5) corresponding to the center of the 2D-electron cloud (see Fig.~\ref{Fig1}).
We note that combined X-ray and neutron powder diffraction measurements on deuterated Y$_2$CD$_x$ with $x=2$ and 2.55 (fully deuterated) reveal that deuterium also occupies interlayer space between Y layers~\cite{Y2CD_Maehlen1,Y2CD_Maehlen2}.
While crystal structures of the deuteride are different from the present compound, one of the deuterium site reported for Y$_2$CD$_{2.55}$ corresponds to the muon site described above.
\par
Since the muon-electron hyperfine interaction is predominantly determined by magnetic dipolar interaction, the hyperfine parameter is calculated using the dipolar tensor ${\bf A}^{\rm dip}$ for the muon site (see Supplemental Material \cite{suppl} for more details).
For the better evaluation of ${\bf A}^{\rm dip}$ which depends on the local atomic configuration, we calculated structural relaxation around the muon site using the OpenMX code, which is based on the DFT+GGA and norm-conserving pseudo potential method~\cite{openmx}.
Assuming 1$\mu_{\rm B}$ at the Y site, the magnitude of hyperfine parameter effective for the shift was approximately evaluated by the root mean square of the diagonal elements of ${\bf A}^{\rm dip}$ to yield $\left|A_{\rm av}^{\rm dip}\right|\simeq73.3$ [mT$/\mu_{\rm B}]$ \cite{suppl}.
A comparison with the experimentally obtained value yields the moment size at the Y site to be, $|A_\mu|/\left|A_{\rm av}^{\rm dip}\right|\simeq0.718(4)\mu_{\rm B}$, which is in line with $\mu_{\rm eff}$ estimated from the measurement of uniform susceptibility for the pc sample (see table~\ref{table}).
We also made a more realistic evaluation based on the powder pattern~\cite{powder,suppl},
where the hyperfine parameter $A^{\rm dip}_{\rm pwdr}$ is derived from a principal value of diagonalized ${\bf A}_{\rm dip}$, yielding $-$27.18~[mT/$\mu_{\rm B}$].
  The comparison with the observed hyperfine parameter ($A_\mu$) yields the Y moment size of 1.94(1)$\mu_{\rm B}$, which is considerably greater than that estimated from the Curie-Weiss analysis.
While this inconsistency may suggest the possibility of other muon stopping site(s),
the conclusion is unchanged that the Curie-Weiss behavior in the pc sample is due to the localized moments of the 4$d$ electrons at the Y site.

The DFT calculation has shown that ferromagnetic instability (Stoner type) in Y$_2$C is indeed suppressed when considering the onsite Coulomb energy ($U$) for the 4$d$ orbital~\cite{Inoshita_theory}. Therefore, the absence of magnetic order as well as the appearance of localized moments at the Y site in the pc sample strongly suggest the importance of electron correlation in gaining a detailed understanding of the electronic ground state in Y$_2$C.
In fact, ARPES experiments have shown that the observed band structure is better described by DFT calculation where $U$ for the 4$d$ orbital~\cite{Horiba_ARPES} is taken into account.
However, we note that our result does not necessarily disfavor the recently suggested scenario that attributes the discrepancy between ARPES and DFT to the surface state stemming from ``topological'' nature of the electride band~\cite{Huang_2018_DFT}.

Considering the case of the sc sample, the temperature dependence curve for $K_\mu$ shown in the right axis of Fig.~\ref{Fig4} indicates nearly no dependence on temperature; here, it is noted that once again, a behavior similar to that of $\chi(T)$ is observed.
This indicates the absence of localized moments in our sc sample, which is qualitatively in line with the phenomenon of Pauli paramagnetism expected in conventional metals. The totally contrasting result obtained for the pc sample suggests that the electronic ground state of Y$_2$C is at the limit between weak-and-strong electronic correlation, where effective shielding of the onsite Coulomb repulsion between the 4$d$ electrons in the electride band may be sensitive to local modulation or to a shift of the Fermi level due to defects/impurities. A detailed impurity analysis indicates that the sample can be unintentionally doped with tungsten during melting process~\cite{private} and can also contain a few percent of carbon deficiency that may lead to electron doping~\cite{Otani_synthesis}.
It is reasonably speculated that such difficulty in controlling impurities may be the background for the prevailing sample dependence.
\par

Remarkably, $K_\mu$ in the sc sample exhibits a negative value ($-40$ to $-50$~ppm).
It should be noted that $K_\mu$ for the conventional metals exhibits a positive shift, which is proportional to the density of states at the Fermi surface; the shift due to $s$ electron is given by $$K_s=\frac{8\pi}{3}\left|\phi_{k_{\rm F}}(0)\right|^2\chi_{\rm p}>0,$$ where $\chi_{\rm p}$ is the Pauli paramagnetic susceptibility and $\phi_{k_{\rm F}}(0)$ is the amplitude of the $s$-electron wave function at the muon position. Considering the observation that the negative Knight shift in metals with positive bulk susceptibility is limited to metals such as Ni, Pd, Pt and those exhibiting strong $s$-$d$ hybridization~\cite{Schenk_shift, JASeitchik_Pd_NMR,Segransan_NMR_Ni,FNGygax_Ni,FNGygax_Pt}, the observed result for the sc sample may be related to the hybridization of the 4$d$ and the electride bands.
Another possibility would be the presence of orbital diamagnetism specific to the 2D electride band, the microscopic details of which are yet to be clarified. In any case, more work is needed to achieve a clearer understanding of the anomalous negative Knight shift in diamagnetic Y$_2$C.

In summary, our $\mu$SR study on pc- and sc-Y$_2$C revealed that both samples showed no sign of static magnetic order at temperatures down to 0.024~K.
From muon Knight shift measurements, it was concluded that the emergence/absence of localized moments at the Y sites in both pc- and sc-samples owes its origin to the intrinsic electronic properties of Y$_2$C, which are sensitive to defects and impurities.
Taking into account predictions from DFT calculations, this sensitivity may be attributed to the electronic correlation of the electride band inherited from the Y 4$d$ orbital, which also leads to the suppression of Stoner instability.
Finally, the temperature-independent {\sl negative} Knight shift observed in the single-crystalline sample points to a hitherto unknown contribution of orbital diamagnetism specific to the electride band.
\par
We thank K. Horiba and H. Kumigashira for sharing their ARPES data prior to publication and for helpful discussion. This work was supported by MEXT Elements Strategy Initiative to Form Core Research Center.
We would like to thank the staff of TRIUMF for their technical support during the $\mu$SR experiment.
The muon site calculation was performed under the Large Scale Simulation Program No.14/15-13(FY2014--2015) of the High Energy Accelerator Research Organization (KEK).
HH acknowledges MEXT KAKENHI (Grant No. 17H06153).



\begin{thebibliography}{00}
\bibitem{JLDye_2003} 
J.~L.~Dye, \href{http://science.sciencemag.org/content/301/5633/607}{\Magazine{Science}{301}{607}{2003}}
\bibitem{JLDye_2009} 
J.~L.~Dye, \href{https://pubs.acs.org/doi/abs/10.1021/ar9000857}{\Magazine{Acc. Chem. Res.}{42}{1564}{2009}}
\bibitem{Ichimura_JACS2002} 
A.~S.~Ichimura, J.~L.~Dye, M.~A.~Camblor, and A.~L.~Villaescusa, \href{https://pubs.acs.org/doi/abs/10.1021/ja016554z}{\Magazine{\JACS}{124}{1170}{2002}}
\bibitem{Matsuishi_2003} 
S.~Matsuishi, Y.~Toda, M.~Miyakawa, K.~Hayashi, T.~Kamiya, M.~Hirano, I.~Tanaka, and H.~Hosono, \href{http://www.sciencemag.org/content/301/5633/626.short}{\Magazine{Science}{301}{626}{2003}}
\bibitem{Toda_AdvMater2007} 
Y.~Toda, H.~Yanagi, E.~Ikenaga, J.~J.~Kim, M.~Kobata, S.~Ueda, T.~Kamiya, M.~Hirano, K.~Kobayashi, and H.~Hosono, \href{http://onlinelibrary.wiley.com/doi/10.1002/adma.200700663/full}{\Magazine{Adv. Mater.}{19}{3564}{2007}}
\bibitem{Kim_JACC2007} 
K.-B.~Kim, M.~Kikuchi, M.~Miyakawa, H.~Yanagi, T.~Kamiya, M.~Hirano, and H.~Hosono, \href{http://pubs.acs.org/doi/abs/10.1021/jp072635r}{\Magazine{\JPCC}{111}{8403}{2007}}
\bibitem{Buchammagari_OrgLett2007} 
H.~Buchammagari, Y.~Toda, M.~Hirano, H.~Hosono, D.~Takeuchi, and K.~Osakada, \href{http://pubs.acs.org/doi/abs/10.1021/ol701885p}{\Magazine{Org. Lett.}{9}{4287}{2007}}
\bibitem{Ruszak_CatalLett2008} 
M.~Ruszak, M.~Inger, S.~Witkowski, M.~Wilk, A.~Kotarba, and Z.~Sojka, \href{http://link.springer.com/article/10.1007\%2Fs10562-008-9619-x}{\Magazine{Catal. Lett.}{126}{72}{2008}}
\bibitem{Adachi_MaterSciEngB2009} 
Y.~Adachi, S.-W.~Kim, T.~Kamiya, and H.~Hosono, \href{http://www.sciencedirect.com/science/article/pii/S0921510708005916#}{\Magazine{Mater. Sci. Eng. B}{161}{76}{2009}}
\bibitem{Kitano_NatChem2012} 
M.~Kitano, Y.~Inoue, Y.~Yamazaki, F.~Hayashi, S.~Kanbara, S.~Matsuishi, T.~Yokoyama, S.-W.~Kim, M.~Hara, and H.~Hosono, \href{http://www.nature.com/nchem/journal/v4/n11/full/nchem.1476.html?WT.ec_id=NCHEM-201211}{\Magazine{Nat. Chem.}{4}{934}{2012}}
\bibitem{Toda_NatCommun2013} 
Y.~Toda, H.~Hirayama, N.~Kuganathan, A.~Torrisi, P.~V.~Sushko, and H.~Hosono, \href{http://www.nature.com/ncomms/2013/130829/ncomms3378/full/ncomms3378.html#close}{\Magazine{Nat. Commun.}{4}{2378}{2013}}
\bibitem{Kitano_NatCommun2015} 
M.~Kitano, S.~Kanbara, Y.~Inoue, N.~Kuganathan, P.~V.~Sushko, T.~Yokoyama, M.~Hara, and H.~Hosono, \href{http://www.nature.com/ncomms/2015/150330/ncomms7731/full/ncomms7731.html}{\Magazine{Nat. Commun.}{6}{6731}{2015}}
\bibitem{Fang_2000} 
C.~M.~Fang, G.~A.~de~Wijs, R.~A.~de~Groot, H.~T.~Hintzen, and G.~de~With, \href{http://pubs.acs.org/doi/abs/10.1021/cm0010102}{\Magazine{G. Chem. Mater.}{1847}{12}{2000}}
\bibitem{Lee_Ca2N} 
K.~Lee,	S.~W.~Kim, Y.~Toda, S.~Matsuishi, and H.~Hosono, \href{http://www.nature.com/nature/journal/v494/n7437/full/nature11812.html}{\Magazine{Nature}{494}{336}{2013}}
\bibitem{Oh_2016} 
J.~S.~Oh, C.-J.~Kang, Y.~J.~Kim, S.~Sinn, M.~Han, Y.~J.~Chang, B.-G.~Park, S.~W.~Kim, B.~I.~Min, H.-D.~Kim, and T.~W.~Noh, \href{http://pubs.acs.org/doi/abs/10.1021/jacs.5b12668}{\Magazine{\JACS}{138}{2496}{2016}}
\bibitem{Inoshita_PRX}
T.~Inoshita, S.~Jeong, N.~Hamada, and H.~Hosono, \href{https://journals.aps.org/prx/abstract/10.1103/PhysRevX.4.031023}{\Magazine{\PRX}{4}{031023}{2014}}
\bibitem{Zhang_Y2C} 
X.~Zhang, Z.~Xiao, H.~Lei, Y.~Toda, S.~Matsuishi, T.~Kamiya, S.~Ueda, and H.~Hosono, \href{http://pubs.acs.org/doi/abs/10.1021/cm503512h}{\Magazine{Chem. Mater.}{26}{6638}{2014}}
\bibitem{Inoshita_theory} 
T.~Inoshita, N.~Hamada, and H.~Hosono, \href{https://journals.aps.org/prb/abstract/10.1103/PhysRevB.92.201109}{\Magazine{\PRB}{92}{201109(R)}{2015}}
\bibitem{Horiba_ARPES}
K.~Horiba, R.~Yukawa, T.~Mitsuhashi, M.~Kitamura, T.~Inoshita, N.~Hamada, S.~Otani, N.~Ohashi, S.~Maki, J.~Yamaura, H.~Hosono, Y.~Murakami, and H.~Kumigashira, \href{https://journals.aps.org/prb/abstract/10.1103/PhysRevB.96.045101}{\Magazine{\PRB}{96}{045101}{2017}}
\bibitem{Otani_synthesis} 
S.~Otani, K.~Hirata, Y.~Adachi, and N.~Ohashi, \href{http://www.sciencedirect.com/science/article/pii/S0022024816304614}{\Magazine{J. Cryst. Growth}{454}{15}{2016}}
\bibitem{Park_2017JACS} 
J.~Park, K.~Lee, S.~Y.~Lee,, C.~N.~Nandadasa, S.~Kim, K.~H.~Lee, Y.~H.~Lee, H.~Hosono, S.-G.~Kim, and S.~W.~Kim, \href{https://pubs.acs.org/doi/10.1021/jacs.6b11950}{\Magazine{\JACS}{139}{615}{2017}}
\bibitem{VASP} 
G.~Kresse, and J.~Hafner, \href{http://journals.aps.org/prb/abstract/10.1103/PhysRevB.47.558}{\Magazine{\PRB}{47}{558}{1993}}
\bibitem{Y2CD_Maehlen1}
J.~P.~Maehlen, V.~A.~Yartys, and B.~C.~Hauback, \href{https://www.sciencedirect.com/science/article/pii/S0925838802010289}{\Magazine{J. Alloys Compd.}{351}{151}{2003}}
\bibitem{Y2CD_Maehlen2}
J.~P.~Maehlen, V.~A.~Yartys, and B.~C.~Hauback, \href{https://www.sciencedirect.com/science/article/pii/S0925838802012458}{\Magazine{J. Alloys Compd.}{356-357}{475}{2003}}
\bibitem{suppl}
See Supplemental Material at http://link.aps/org/supplemental/``DOI-TBA''
\bibitem{openmx}
T.~Ozaki, \href{https://journals.aps.org/prb/abstract/10.1103/PhysRevB.67.155108}{\Magazine{\PRB}{67}{155108}{2003}} http://www.openmx-square.org
\bibitem{powder} 
See, for example, C.~P.~Slichter, {\it Principles of Magnetic Resonance}, Springer, New York, 1990 3rd ed.

\bibitem{Huang_2018_DFT} 
H.~Huang, K.-H.~Jin, S.~Zhang, and F.~Liu, \href{https://pubs.acs.org/doi/abs/10.1021/acs.nanolett.7b05386}{\Magazine{Nano Lett.}{18}{1972}{2018}}
\bibitem{private}
K.~Ohashi, private communication.
\bibitem{Schenk_shift}
A.~Schenck, {\it Muon Spin Rotation Spectroscopy: Principle and Applications in Solid State Physics}: Adam Hilger, Bristol, 1985, p.128
\bibitem{JASeitchik_Pd_NMR}
J.~A.~Seitchik, A.~C.~Gossard, and V.~Jaccarino, \href{https://journals.aps.org/pr/abstract/10.1103/PhysRev.136.A1119}{\Magazine{Phys. Rev}{136}{A1119}{1964}}
\bibitem{Segransan_NMR_Ni}
P.~J.~Segransan, Y.~Chabre, and W.~G.~Clark, \href{http://iopscience.iop.org/article/10.1088/0305-4608/8/7/024/meta}{\Magazine{J. Phys. F: Met. Phys.}{8}{1513}{1978}}
\bibitem{FNGygax_Ni}
F.~N.~Gygax, W.~R${\rm \ddot u}$egg, A.~Schenck, H.~Schilling, W.~ Studer, and R.~Schulze, \href{http://www.sciencedirect.com/science/article/pii/0304885380902498}{Proc. ICM 79, J.M.M.M., {15--18}, M${\rm \ddot u}$nchen 1979 (1980), p. 1191}
\bibitem{FNGygax_Pt}
F.~N.~Gygax, A.~Hintermann, W.~R${\rm \ddot u}$egg, A.~Schenck, W.~Studer, A.~J.~van~der~Wal, and L.~Schlapbach, \href{https://link.springer.com/article/10.1007%2FBF02065929}{\Magazine{Hyperfine Interaction}{17}{377}{1984}}
\end{thebibliography}
\end{document}